\documentclass[preprint,12pt]{elsarticle}

\usepackage{amsmath}
\usepackage[english]{babel}
\usepackage{graphicx}

\journal{Physica A}

\begin{document}

\begin{frontmatter}

\title{Robust linear regression with broad distributions of errors}

\author{Eugene B. Postnikov}
\ead{postnicov@gmail.com}
\address{Theoretical Physics Department, Kursk State University, Radishcheva st., 33 Kursk 305000, Russia}

\author{Igor M. Sokolov}
\ead{igor.sokolov@physik.hu-berlin.de}
\address{Institute of Physics, Humboldt-University at Berlin,
Newtonstr.15, 12489, Berlin, Germany}

\begin{abstract}
We consider the problem of linear fitting of noisy data in the case of broad (say $\alpha$-stable) distributions of random impacts (``noise''), which can lack even the first moment. 
This situation, common in statistical physics of small systems, in Earth sciences, in network science or in econophysics, does not allow for application of conventional Gaussian maximum-likelihood estimators 
resulting in usual least-squares fits. Such fits lead to large deviations of fitted parameters from their true values due to the presence of outliers. 
The approaches discussed here aim onto the minimization of the width of the distribution of residua. The corresponding width of the distribution can either be defined via the 
interquantile distance of the corresponding distributions or via the scale parameter in its characteristic function. 
The methods provide the robust regression even in the case of short samples with large outliers, and are equivalent to the 
normal least squares fit for the Gaussian noises. Our discussion is illustrated by numerical examples. 

{\it \bf Highlights:}

\begin{itemize}
\item Correct estimating of the linear fit parameters in a presence of large outliers

\item The median of the empirical distribution of the residues determines line's shift

\item The minimum of interquantile width determines line's slope (1st method)

\item The maximum of characteristic function's residues determines line's slope (2nd method)
\end{itemize}

\end{abstract}

\begin{keyword}
L\'evy noise \sep data processing \sep linear fit
\end{keyword}

\end{frontmatter}

\section{Introduction}

The method of least squares linear regression (straight line fitting) has a very long history: it was invented in its simplest form by C.F.~Gau\ss, but is still one of the most widespread and powerful approaches in data analysis. 
It may be used as a stand-alone tool to detect linear trends, or be incorporated into more complex analysis procedures, like Detrended Fluctuation Analysis proposed in \cite{Peng1994},
whose first step requires subtraction of linear trends from subpartitions of data. The standard variant of the method assumes the linear relation between the dependent variable $y$ and the independent one $x$, 
and the existence of a random impacts on the outcomes of single measurements, represented by the noise $\xi$, so that 
\begin{equation}
 y_i = ax_i + b + \xi_i,
 \label{LiRe1}
\end{equation}
and is aimed onto extracting information about $a$ and $b$ from such noisy data. 
The standard method works well if the data are ``compact'', i.e. when the corresponding interval on the abscissa is homogeneously
sampled and no large ordinate outliers are present. The method is essentially a parametric one and can be regarded as the maximum likelihood approach assuming the Gaussian distribution of independent errors. 
The challenges of more complicated samples originating from modern problems of experimental and computational physics and related fields
have motivated works aimed to improve the accuracy of fits to extremely irregular data, i.e. the ones having outliers on the ordinate and on the abscissa (leverage points), or large
errors in locating $x_i$, see \cite{Macdonald1992,Cantrell2008} for the list of modern modifications. For this reason, a number of works discuss the criteria for a detection of this outliers with the following their elimination with respect to a prescribed cut-off level, and the regression of obtained ``cleared'' samples \cite{Rousseeuw2005} or a choice of subintervals, where the influence of outliers 
could be negligible \cite{Grech2013,Gulich2014}. Another problem arises for the non-independent noises which themselves can show trends \cite{Hu2001}. 

Even in the case of independent errors the problems arise if the noise possess a heavy-tailed distribution, i.e. generates large outliers. 
These are quite characteristic for a large variety of process in small nonequilibrium systems, network dynamics, econophysics, etc. \cite{Clauset2009}. 
Since these distributions may lack even the first moment, their processing, if keeping the principles of the least-square regression untouched,
requires very specific methods \cite{Adler,Gather2006} including repeated median regression, the consideration of a nested hierarchy block subdivisions for the analyzed sample, etc. 
For such cases non-parametric regression methods may be superior to the standard one. 

In the present work we discuss two such
approaches, the quantile regression as pioneered by Koenker and Basset \cite{Koenker1978}, and the scale parameter regression based on the 
properties of characteristic functions. The methods are non-parametric (i.e. do not assume the specific form of the distribution) and robust (i.e. do not rely
on the existence of its moments). Our numerical examples consider linear trend in presence independent errors distributed according to L\'evy stable laws.

As a practical example, we consider geophysical data, namely the eastward component of the geomagnetic field measured on a moving Antarctic ice shelf, showing a linear trend from the motion and a combination of small and large scale fluctuations. Here the results of robust scale parameter regression are compared to 
conventional methods. 

\section{Linear regression}

Before discussing the specific methods, let us shortly review the general idea (or, better, general ideas) of linear regression.
Posing the regression problem starts from the assumption that the values of the dependent variable (observable) $y_i$ linearly depend on  
$x_i$, but are subject to additive noise $\xi_i$, Eq.(\ref{LiRe1}).
We are looking for the way of inferring of the parameters $a$ and $b$, delivering the best possible estimates $\hat{a}$ and $\hat{b}$
for these parameters. 
In the ideal situation (at least in the asymptotic setting when the total number of measurement points
$N$ gets large, $N \to \infty$) the method should give $\hat{a}=a$, $\hat{b}=b$. In praxis, this is usually done by the application of the least squares fit. 

There are different ways to think about the least squares method. 

First, we can follow the standard line of argumentation pertinent to statistical inference and make a maximal likelihood estimate for the 
parameters $a$ and $b$ assuming the distribution of $\xi_i$ is Gaussian with zero mean and unknown dispersion, 
\[
 p(\xi_i) = \frac{1}{\sqrt{2 \pi} \sigma} \exp \left( - \frac{\xi_i^2}{2 \sigma^2} \right).
\]
In this case the probability density of a given realization of $\xi_i$ is given by the product of such single-point distributions:
\[
 p(\xi_1,...,\xi_N) = \prod_{i=1}^N p(\xi_i) = \left(\sqrt{2 \pi} \sigma \right)^{-N}  \exp \left( - \frac{\sum_{i=1}^N \xi_i^2}{2 \sigma^2} \right).
\]
Changing from $\xi_i$ to $y_i$ we get the corresponding density of the experimental outcomes $\{y_i\}$,
\[
 p(y_1,...,y_N | a,b) = \left(\sqrt{2 \pi} \sigma \right)^{-N}  \exp \left( - \frac{\sum_{i=1}^N (y_i - a x_i -b)^2}{2 \sigma^2} \right).
\]
Considering the log-likelihood of $a$ and $b$ provided the data,
\[
 L(a,b | \{y_i\}) = \ln p(\{y_i\} | a,b) = const - \frac{\sum_{i=1}^N (y_i - a x_i -b)^2}{2 \sigma^2}
\]
and maximizing it with respect to $a$ and $b$, we get the least square prescription for finding $a$ and $b$ by minimizing the sum of squared residues
\[
R^2=\sum_i \left[y_i-(a x_i+ b)\right]^2 = min.
\]
Note that this criterion, which essentially assumes the Gaussian prior is of course a parametric one, and therefore not robust. 
Assuming another distribution, say the Laplace one with 
\[
 p(\xi_i) = \frac{1}{\sigma} \exp \left( - \frac{|\xi_i|}{\sigma} \right)
\]
will lead to a different criterion, in this case to the minimization problem of 
\[
 R = \sum_i \left|y_i-(a x_i+b)\right|.
\]

Another approach to the linear regression is a geometric one. As above, 
the variables $\xi_i$ are assumed to be i.i.d. random variables drawn from a distribution $p_\xi(\xi)$, which we will
be assumed continuous, symmetric and monomodal. The coordinates of points $(x_i,\xi_i)$ 
are mutually independent. The points $(x_i,\xi_i)$ are considered as realizations of points in a two-dimensional \textit{cloud}
characterized by the density (joint probability density) $p(x,\xi) = p(x) p(\xi)$. This cloud is mirror-symmetric with respect to $x$ axis.
The pairs $(x_i,y_i)$ with $y_i$ depending on $x_i$ are realizations of points of another two-dimensional cloud,
which is obtained from the first one by a shift and an affine transformation. The regression aims on the restoration of these
transformation parameters $a$ and $b$ so that the cloud with the density $p(x, \xi)$ with $\xi = y -(ax +b)$ indeed has the properties discussed above. 
One looks for the empirical estimators $\hat{a}$ and $\hat{b}$ of these parameters.

If we say that this symmetry presumes the fact that the center of mass of the cloud lays on $x$
axis and then that one of its main axes of inertia coincide with it, we get from the first requirement
\[
 \sum_i y_i-(ax_i+b) = 0
\]
so that $b = N^{-1} \sum_i (y_i-ax_i) = \langle y \rangle - a \langle x \rangle$. Then one notes that the main axes of inertia of the
two-dimensional body are such that the moments of inertia with respect to these are extremal, and requires the extremality of
\[
 I = \sum_i [y_i-(ax_i+b)]^2 = \sum_i [(y_i - \langle y \rangle - a (x_i - \langle x \rangle)]^2
\]
(with $I$ being the moment of inertia with respect to the $x$-axis) with respect to $a$  with $b$ defined as before.
This gives equations which define $a$ and $b$ from the least square method. 
However, the mirror-symmetry of the $(x, \xi)$-cloud with respect to $x$-axes can be cast into different other extremality prescriptions
or into the statement that half of its mass has to lay above, and half below the axis, which gives (provided $b$ is defined) the robust median
regression for $a$. The method should work in this form provided $\langle y \rangle$ and $\langle x \rangle$ do exist. If they do not 
(i.e. when the distribution of $y$ is broad (outliers) or the distribution of $x$ is broad (leverage points)), the standard problems
arise. Note 
that the median method is sensitive to the centering of the cloud: it will break down if the center of the cloud is at the origin.

Another variant of the geometric approach discussed below is based on a different consideration. It aims on finding the
estimate for $a$ prior to connecting it to $b$ and is robust both with respect to outliers and to leverage points (which question is not a 
topic of the present work). 

Let us define the residues 
\[
\Delta y_i = (\hat{a}-a)x_i + (\hat{b}-b) + \xi_i,
\]
and concentrate first on the obtaining of the best estimate $\hat{a}$ for the slope parameter $a$. 
We note that the parameter $b$ only shifts the distribution of $\Delta y_i$, and only influences the position of $\Delta y_i$, while the parameter
$a$ influences the width of $p(\Delta y_i)$. In the case of exact tuning $\hat{a} = a$ this width is given by the one 
of the distribution of $\xi$; for $\hat{a} \neq a$ the distribution of $\Delta y_i$ (centered on $\hat{b}-b$)
is a \textit{convolution} of the distribution of $\xi$ and the one of $(\hat{a}-a)x_i$, which now has a nonzero width.
Since the convolution of two distributions is always "broader" than each of them, the minimal width will coincide with 
the one of the distribution of $\xi$ and achieved for $\hat{a} = a$. In a setting when the width of the distribution is 
given by its variance, the method again reduces to the least squares approximation: The empirical width is defined as
\[
 W^2 = \frac{1}{N} \sum_i \Delta y_i^2  
\]
and is minimized with respect to two free parameters $\hat{a}$ and $\hat{b}$. 

\section{Width regression}

In our approach we use the fact that while the parameter $b$ only shifts the distribution of $\Delta y_i$, and influences the position of $\Delta y_i$,
the width of the distribution of $\Delta y_i$ is only influenced by the parameter $a$. 
In the case of exact tuning $\hat{a} = a$ this width is given by the one 
of the distribution of $\xi$. 
Our two regression approaches differ in the
point of how this ``width'' of the distribution is defined.

As we have already seen,  defining the width by a variance of the corresponding distribution (provided it exists) leads to the standard least
square prescription; its additional advantage is that the minimization procedure follows by solution of a system of linear 
algebraic equations. Other definitions of width (for example estimation the first absolute moment of the distribution) 
lead to nonlinear equations which have to be solved numerically. Both methods estimate
width via some absolute moments of the distribution. Both methods do not work for distributions having
power-law tails; the first one fails for the ones with diverging second moment, the second one for the distributions
with diverging first moment (like Cauchy distribution). 

Moments do not represent robust statistics since they do not exist for all distributions. The robust statistics is given by such
measures of width which exist for all distributions of $y$ and of $x$. There are several classes of such robust measures
either pertinent to the distribution itself, say, its quantiles, or to its characteristic function, say its scale parameter.
These two possibilities will be discussed in the forthcoming sections. In all our discussions we will only concentrate on outliers, and
both in our numerical examples and in the practical one $x_i$ are homogeneously distributed within a finite interval.

\subsection{The interquantile distance regression}

\subsubsection{Description of the method}

One of the robust estimates of width is given by interquantile distance of the corresponding distribution (since the cumulative
distribution function ({\it c.d.f.}) and therefore the quantiles do exist for any proper PDF). 

The practical realization for a given set of data $\{x_j\}$, $j=1..N$ is subdivided into two steps.
Since the width of {\it c.d.f.} is invariant with respect to shifts, at the first step we consider the series 
$$
y^{i}_j=y_j-a_ix_j
$$
and its {\it c.d.f.}'s $C(a_i)$ for the trial slopes $a_i$ equidistantly sampled with the step $h_a$ within some interval.
We moreover fix some quantiles $q$ and $1-q$ defining the width to be minimized (in the following examples we set $q=1/4$). 
As it has been discussed above, the minimal half-width of $C(\hat{a})$ corresponds to the best fit of $a_i=\hat{a}$. 

For each $a_i$, the obtained set of values $y^{i}_j$ is sorted in ascending order to $\tilde{y}^{i}_j$, whence the desired {\it c.d.f.} half-width is simply
\begin{equation}
HW(C(a_i))=\tilde{y}^{i}_{[3N/4]}-\tilde{y}^{i}_{[N/4]}.
\label{hwk}
\end{equation}
A search of the minimum for the series (\ref{hwk}) provides the index of desirable value $a_i=\hat{a}$. Here the square brackets denote an integer part of the fractions.
Having obtained $\hat{a}$, one can obtain the shift parameter $\hat{b}$ as the median of the distribution of $y_j-\hat{a} x_i$. 
However, it should be pointed out that it might be preferable to obtain the value of $\hat{b}$ via the equidistant trials $b_i$, for which the {\it c.d.f.} of the series
$$
y^{i}_j=y_j-\hat{a}x_i-b_i
$$
has a median equal to zero instead of the single-run median search. This is the case for non-equispaced samples, since the algorithms for identifying the zero crossing
provide better accuracy due to the possibility of interpolation.

Practically, due to sample's discreteness, we use the criterion of minimum for $|\tilde{y}^{j}_{[N/2]}|$, where $\tilde{y}^{i}_j$ is again the series of ${y}^{i}_j$ sorted in ascending order.
Thus both fitted parameters, $a$ and $b$ are determined.

Although in our simple realization of the method we mostly obtain the quantiles by simply counting the points, we note that this can be done in a
more elegant way using the quantile regression methods as pioneered in \cite{Koenker1978} (see \cite{Koenker} for the state of the art discussion). 
This general approach can be cast into the minimization problem, namely, in solving 
\begin{equation}
\hat{a}=\mathrm{arg min}_{a\in\Re}\sum_{i=1;\, y_j\geq ax_i}^{n}q|y_i-ax_i|+\sum_{i=1;\, y_j< ax_i}^{n}(1-q)|y_i-ax_i|,
\label{minsolv}
\end{equation}
where $0<q<1$ is the regression quantile sought for. Formally, the method requires the existence of the first moment of the $y$-distribution, and may 
lead to instabilities when applied to situations with large outliers, although we never encountered them is our test runs.

\subsubsection{Maximizing sensitivity}

It should be pointed out that although the approach works for an arbitrary part of {\it c.d.f}'s width, the important question is, what quantile has to be chosen 
to provide the largest local sensitivity of the method. 

Let us at the beginning consider a centered distribution and take $\hat{b}-b = 0$. Let us denote $\Delta a =\hat{a}-a$. 
The distribution of centered $y$ is a convolution of the distributions of $(\hat{a}-a)x$ and of $\xi$, since $\xi_i$ are
independent on $x$.

For $x$ homogeneously distributed on the interval $[-W/2, W/2]$ the convolution
$\tilde{p}(y)$ of the corresponding distributions can be expressed via the cumulative distribution function
$C(x)= \int_{-\infty}^x p_\xi(\xi)d\xi$, namely
\begin{equation}
\tilde{p}(y) = \frac{1}{W\Delta a} \left[C \left(y+\frac{W \Delta a}{2} \right) - C \left(y - \frac{W \Delta a}{2} \right) \right].
\label{eqpfull}
\end{equation}
For $\Delta a$ very large the distribution tends to a rectangular of width $W\Delta a$, so that its interquantile distance 
(for given quantiles of index $q$ and $1-q$) is linear in $\Delta a$. For $\Delta a$ small the dependence of
interquantile distance on $\Delta a$ gets quadratic.

Let us discuss this situation by expanding the cumulative functions $C$ in Eq.~(\ref{eqpfull}) in Taylor series around $y$.
Since all even terms vanish, only the terms linear and cubic in $W \Delta a/2$ survive in the lowest orders, so that
\begin{equation}
\tilde{p}(y) = \frac{1}{W\Delta a} \left[C'(y) W \Delta a + C'''(y) \frac{W^3 \Delta a^3}{3} +... \right] = p(y) + \frac{p''(y)}{3}W^2 \Delta a^2.
\label{taylor}
\end{equation}
The position $Q_q$ of the $q$-th quantile is given by
\begin{equation}
\int_{-\infty }^{Q_q} \tilde{p}(y) dy = q.
\label{quant}
\end{equation}
Inserting the expression Eq.(\ref{taylor}) into Eq.(\ref{quant}) and performing the integration we get $Q_q$ as the solution of the equation
\[
C(Q_q)+ \frac{W^2 \Delta a^2}{3} p'(Q_q) = q.
\]
We note that for $\Delta a = 0$ the solution of $C(Q_q) = q$ gives exactly the quantile of the distribution of the noise,
so that
\[
C(Q_q) - q = p(Q_q) \Delta Q_q
\]
is proportional to the shift of this quantile when detuning $a$. The highest sensitivity is attained when the largest
absolute shift $|\Delta Q_q|$ for given $\Delta a$ is observed.
Since 
\[
\Delta Q_q = - \frac{W^2}{3} \frac{p'(Q_q)}{p(Q_q)} \Delta a^2,
\]
this takes place when $q$ is chosen such that the absolute value of the logarithmic derivative 
\[
\left| \frac{p'(Q_q)}{p(Q_q)} \right| = \mathrm{max}
\]
is attained at the point $Q_q$ of the error distribution. For example for the Cauchy distribution this are
\textit{exactly} the lower and the upper \textit{quartiles} of the distribution.   

For a Gaussian distribution, for which the logarithmic derivative equals to $Q_q$ the absolute relative change in the quantile
\[
\left| \frac{\Delta Q_q}{Q_q} \right| = \frac{W^2}{3} \Delta a^2,
\]
doesn't depend on the index. However, it should be kept in mind in practical applications that the chosen quantile must contain a sufficient number of points.

\subsubsection{Numerical example}

Let us consider the signal $y=ax+b+\xi$, where  $\xi$ is a random variable with the symmetric null-centered $\alpha$-stable density  with the characteristic function
\begin{equation}
\phi(\omega)=\exp\left(-\gamma^{\alpha}|\omega|^{\alpha}\right),
\label{pdf}
\end{equation}
where $\alpha\in(0,\,2]$ is the characteristic exponent and $\gamma>0$ is the scale parameter.
Note that for $\alpha<2$ the second moment is absent, therefore the dispersion-based methods are inapplicable, and for $\alpha\in(0,\,1]$ 
even the mean value diverges, thus one can not apply the approaches calculating the absolute values of deviations. 

\begin{figure}
\includegraphics[width=\textwidth]{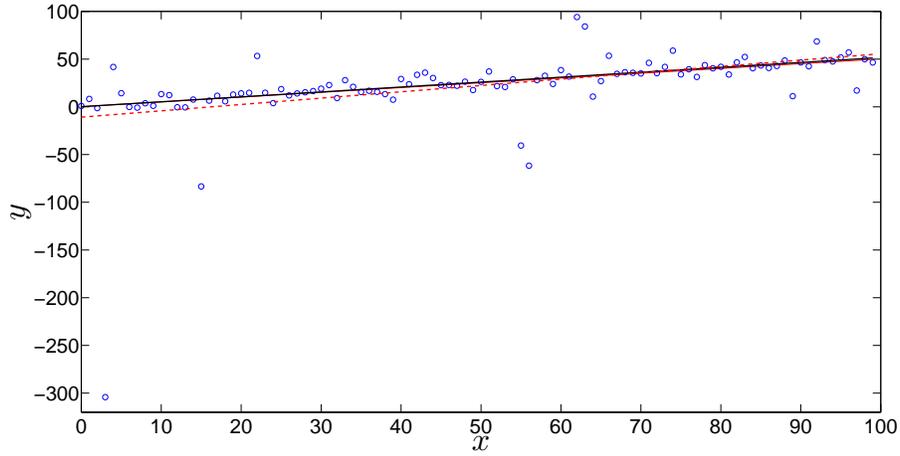}
\caption{The initial deterministic process (thin solid line, almost invisible because it is overlapped by the fit line), its sample with the added 
L\'evy noise (circles) and the results of fitting by the proposed method (thick solid line) and by the conventional least squares method (dashed line).}
\label{example}
\end{figure}

Fig.~\ref{example} demonstrates an example of the fitting for the function $y=0.5x+0.2$ corrupted by the white L\'evy noise with $\alpha=1$ (Cauchy distribution) 
with the scale parameter $\gamma=5$, i.e. with a quite large outliers, over the time interval $t\in[0,\,100]$ sampled with the unit step. 
The random numbers are generated by the routine {\tt stblrnd} \cite{routine} based on the methods presented in \cite{Chambers1976,Weron1995}. 
The sample is processed by the written MATLAB routine with the step size $0.001$ for both $a$ taken from the interval 
$[0,\,1]$ and $b$ taken from $[-10,\,10]$. The obtained pair $(a,b)=(0.515,\,0.149)$, while the conventional least squares method of linear fit provides sufficiently worse values $(0.664,\,-10.807)$.

\begin{figure}
\includegraphics[width=\textwidth]{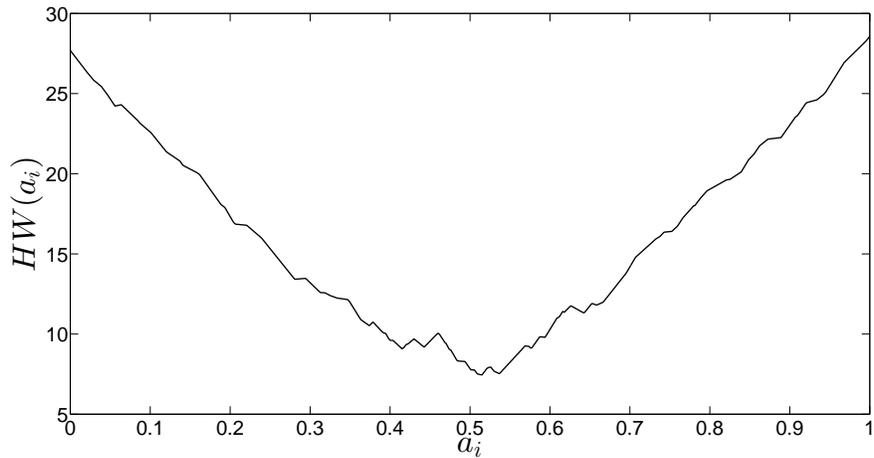}
\caption{The half-width of the cumulative distribution function for the samples $x_j-a_ix_j$ as a function of the trial slopes $a_i$ for the data shown in the Fig.~\ref{example}.}
\label{halfwidth}
\end{figure}

Fig.~\ref{halfwidth} demonstrates the behavior of the basic statistics of the method, the half-width of the cumulative distribution function. It is naturally irregular since a single random realization is processed. 
However, the global minimum is clearly visible even on the background of multiple small local ones. Note that the presence of local minima
might be a problem if the global one is shallow, as it happens in the example of Sec. \ref{Pe}. Therefore it is always advisable to plot the curve like in Fig.~\ref{halfwidth}
to be able to estimate the possible uncertainties caused by this effect. In the case when such uncertainties are large it is better to resort to the method described in the next section.

\begin{figure}
\includegraphics[width=\textwidth]{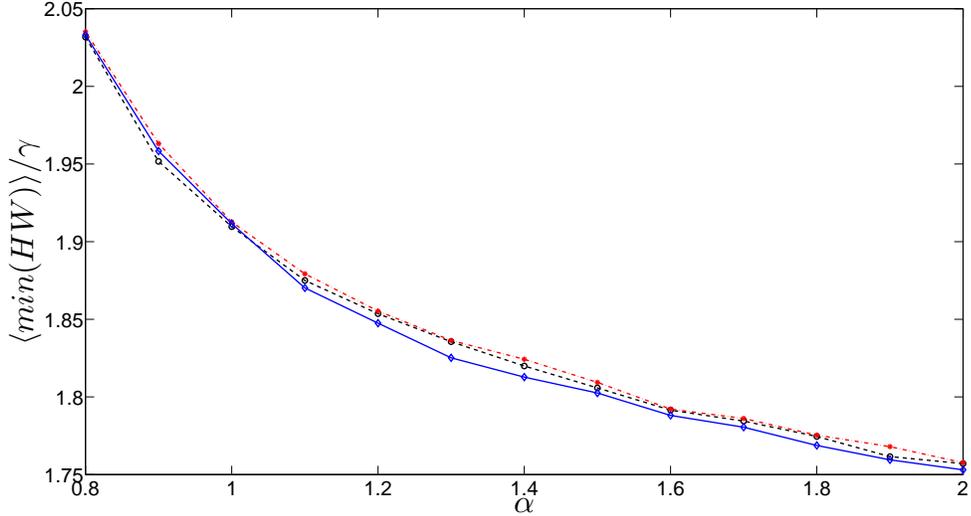}
\caption{The dependence of the minimal width for the cumulative distribution functions normed by the scale coefficients for the various characteristic exponents $\alpha$ and scales $\gamma=0.5$ 
(diamonds connected by the solid line (blue in color online)), $\gamma=1$ (circles connected by the dashed line (black in color online)) and $\gamma=5$ (asterisks connected by the dash-dotted line (red in color online)).}
\label{ensemble}
\end{figure}

Fig.~\ref{ensemble} shows the behavior of scaled half-width of the \textit{c.d.f.} for different characteristic exponents and scale factors, i.e. the $\alpha$-dependence of $HW/\gamma$.
The curves are results of ensemble averaging over 10000 realizations. One can see that they all monotonically decrease when the distribution of errors tends to the normal distribution and  have a 
universal shape (the deviations are within the error of averaging). This fact follows from the self-similarity of the distributions (\ref{pdf}) since their arguments depend on the 
combination $x/\gamma$ if the scale parameter is defined as in Eq.(\ref{pdf}). Additionally, this picture shows that although for different $\alpha$ these are the quantiles with
different indices which are most sensitive to the deviation of 
$a$ from its best value $\hat{a}$, fixing interquantile distance (the half-width of {\it c.d.f} in our case) as a test statistics practically provides a uniform quality of slope's determination.

\subsection{Scale parameter regression}

Another method is based on the estimating width of the distribution via its characteristic function $f(k)=\langle \exp(iky) \rangle$,
which is also an object which does exist for any proper distribution. 

Since the distribution of centered $y$ is a convolution of the distributions of $(\hat{a}-a)x$ and of $\xi$, 
its characteristic function $f_y(k)$ is the product of the characteristic functions of the distributions of $\xi$, $f_\xi(k)$, and of $\Delta a x$, being 
$f_{\Delta a}(k) = \int \exp(ik\Delta a x) p(x) dx = f_x(k \Delta a)$: 
\[
f_y(k) = f_\xi(k) f_x(k \Delta a).
\]
For example, for symmetric Levy noise with scale parameter $\gamma$ and homogeneous distribution of $x$ on $(-W/2,W/2)$ we get
\begin{eqnarray}
f_y(k) &=& \exp(-\gamma^{\alpha} |k|^\alpha) \frac{\sin(W \Delta a k /2)}{W \Delta a k /2} \nonumber \\
&\simeq& 1-\gamma^\alpha |k|^\alpha -\frac{W^2 \Delta a^2}{3} k^2 + ...
\label{fsink}
\end{eqnarray}
(where the prefactor of $k^2$ is simply the dispersion of the distribution of $x$). Thus, fixing some $k$ (small enough so that
the asymptotic expansion close to $k=0$ still works for both distributions of $x$ and of $\xi$), we can look for the
maximum in $\hat{a}$ of $f_y(k)$ which is attained exactly at $\Delta a = 0$. 

Note that for Gaussian distribution of $\xi$ Eq.(\ref{fsink}) for small $k$ reduces to
\[
f_y(k) \simeq 1- (\gamma^2 + \gamma_x^2 \Delta a^2) k^2,
\]
describing a centered distribution with the total dispersion
\[
\gamma_{tot}^2=\gamma^2 + \gamma_x^2 \Delta a^2,
\]
so that minimizing the total width using the small-$k$ approach reduces to the minimizing of the 
dispersion of $y_i$; its approximation by an empirical estimator leads to the least squares method. 
The local sensitivity of the method is always given by $\gamma_x^2 k^2$ so that it can be influenced
by a judicious choice of $k$ which has to be small enough to allow using the quadratic approximation
(it depends e.g. on the higher moments of the $x$-distribution) but not too small to make the sensitivity too
low.  
 
This appropriate value of $k$ for an arbitrary $\alpha$ can be determined by the following reasoning. 
The  function $\sin(W \Delta a k /2)/(W \Delta a k /2)$ in the first line of Eq.(\ref{fsink}) is an oscillating function 
whose two roots closest to the global maximum at $k=0$ are located in
$$
a=\hat{a}\pm\frac{\pi}{kW}.
$$
Therefore, if the value of $\hat{a}$ can be restricted to $\hat{a}\in[-a_{max},\,a_{max}]$ by inspection, 
the frequency parameter can by taken as $k=\pi(a_{max}W)^{-1}$. This results in the location of the main maximum 
within the prescribed interval only. 

Therefore, the operational idea of the method is to calculate the empirical characteristic function 
\begin{equation}
\hat{f}(k | a) = \frac{1}{N} \sum_{j=1}^N \exp\left[ik(y_j-ax_j)\right]
\label{hatf}
\end{equation}
as an approximation for $f_y(k)$ for given $a$ and consider its dependence on $a$ for a fixed $k$ within the range described above.

The shift parameter $b$ is omitted in Eq.~(\ref{hatf}) since it only introduces the phase multiplier
\[
\hat{f}(k | a) \to e^{ikb} \hat{f}(k | a),
\]
which can be eliminated by considering
\[
\phi(k, a) = |\hat{f}(k | a)|
\]
(or alternatively by centering in real space).

\begin{figure}
\includegraphics[width=\textwidth]{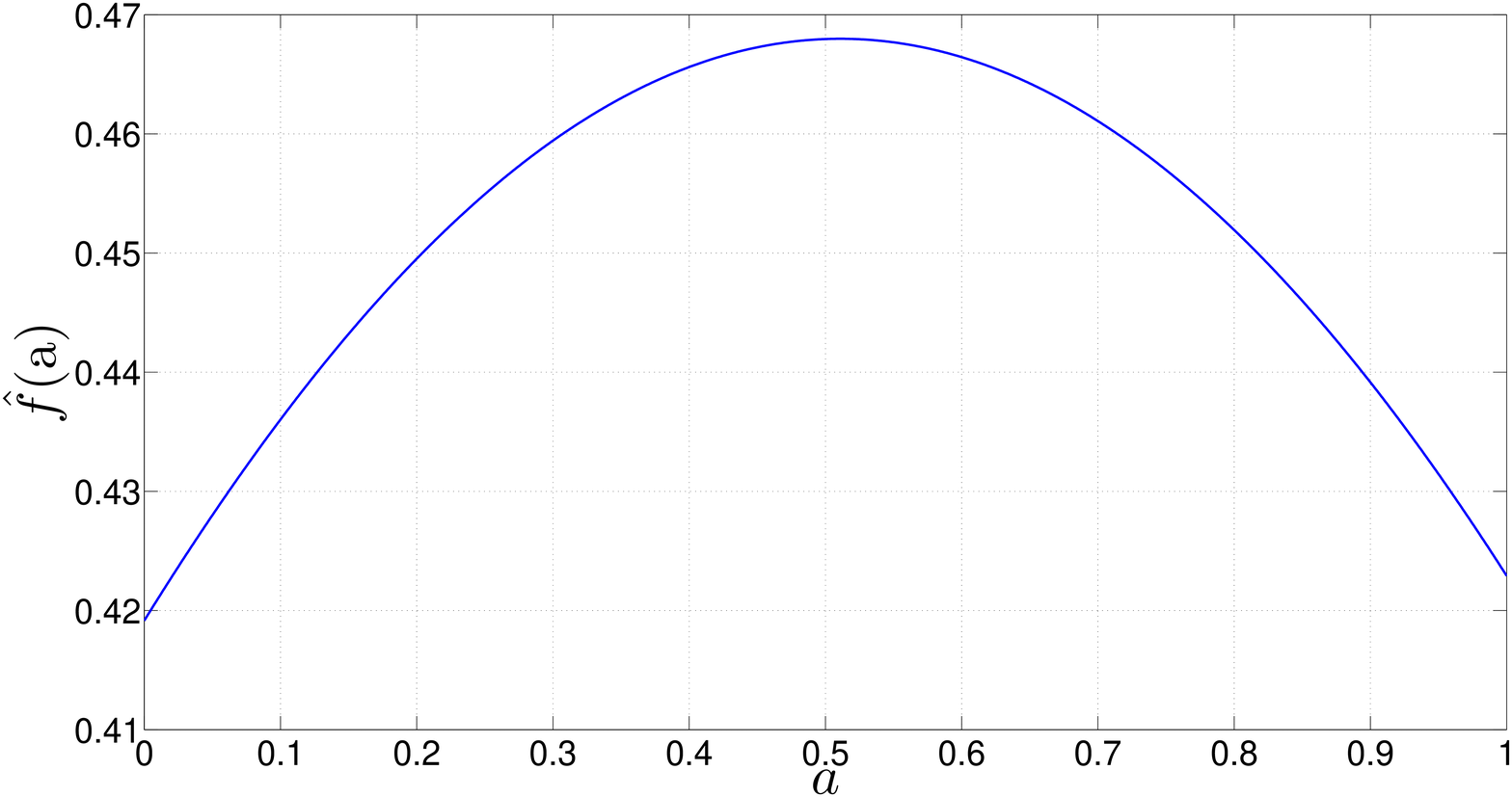}
\caption{The dependence of the characteristic function on the trial slopes around the main maximum. The parameters of the regular and noise components are the same as in Fig.~\ref{halfwidth}.}
\label{kdiffa}
\end{figure}

Fig.~\ref{kdiffa} demonstrates the example of the behavior for the function $\hat{f}(k | a)$ calculated for a single realization of the same linear 
function corrupted by L{\'e}vy noise as in the Subsection 3.2 with the same spacing of the trial parameter $a$. One can clearly see the maximum sought for, 
which allows to determine $\hat{a}=0.504$, a better estimate than the one obtained by the method of the previous section. 
Moreover, the curve is much smoother in comparison with Fig.~\ref{halfwidth} which allows to avoid false extrema.  
The still undefined parameter $b$ can be determined using the median regression of detrended data $y_i-\hat{a} x_i$ as described above, 
since in the present approach it only enters the phase shift and can only be defined modulo $2\pi$.

\subsection{Comparison of the methods}

\begin{figure}
\includegraphics[width=\textwidth]{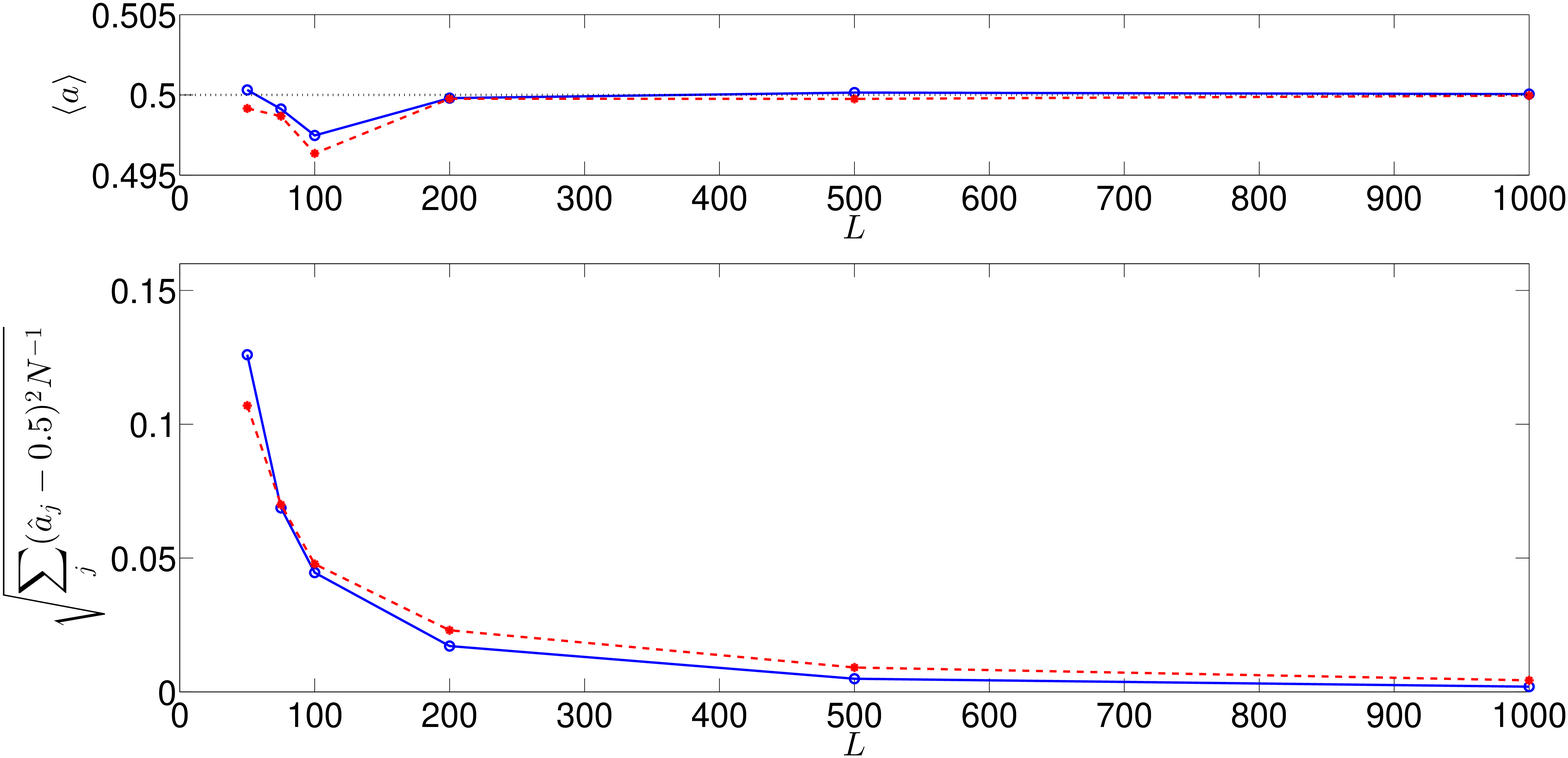}
\caption{Upper panel: the ensemble averaged value of the slope determined via the quantile distribution width method (circles connected by solid lines) and via the characteristic 
function regression (asterisks connected by dashed lines) for various sample lengths. 
Lower panel: the root-mean-square deviations from the exact value for both methods.  The parameters of the regular and noise components are the same in Fig.~\ref{halfwidth}.}
\label{compL}
\end{figure}

\begin{figure}
\includegraphics[width=\textwidth]{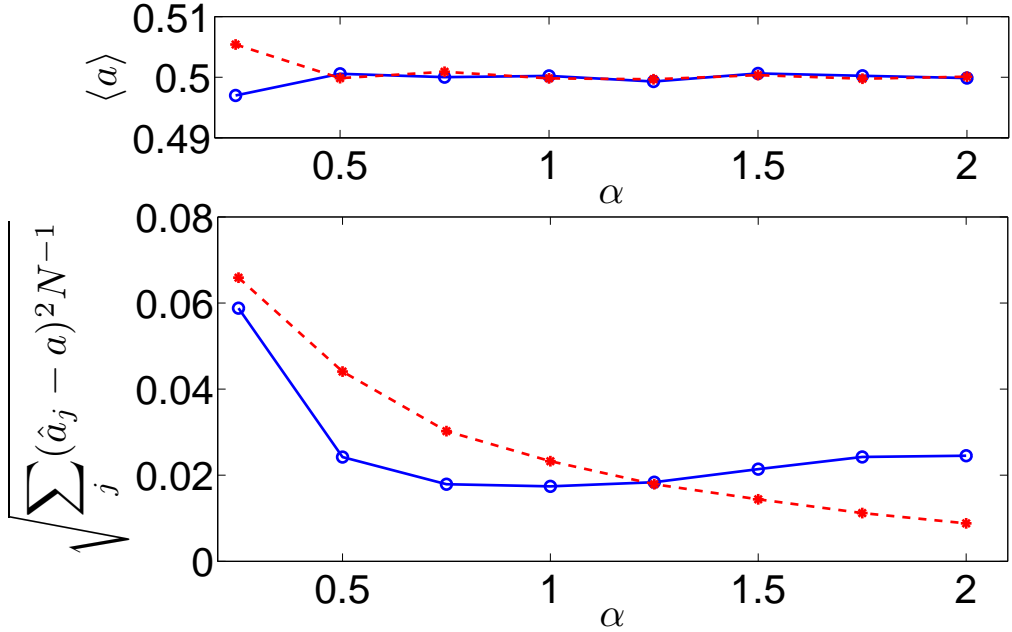}
\caption{Upper panel: the ensemble averaged value of the slope determined via the quantile distribution width method (circles connected by solid lines) and via the characteristic function regression (asterisks connected by dashed lines) for various indices of the noise distribution. Lower panel: the root-mean-square deviations from the exact value for both methods.  The parameters of the regular and noise components are the same in Fig.~\ref{halfwidth}.}
\label{compA}
\end{figure}

Let us compare the efficiency of two proposed methods, primarily in determination of the line's slope. Since individual realizations, especially in the case of small $\alpha$, have a considerable variability, 
we performed the calculations for an ensemble of 1000 individual realizations (with the parameters given above), each of them fitted separately. 
Fig.~\ref{compL} presents the resulting average values of the slope and its root-mean-square deviations from the exact value $a=0.5$.
Fig.~\ref{compA} shows a similar comparison for a fixed sample length ($L=200$) but for different indices $\alpha$ of the noise's distribution.

One can see that both methods provide more than reasonable fitting even for very short samples. The method based on the characteristic function is more accurate for shortest samples that can be explained by the $2\pi$-periodicity of the random phase: since large outliers originated from L{\'evy} noise are relatively rare, their influence in the vicinity of the main frequency maximum is small for short samples while their presence 
in boundary quartiles strongly influences the half-width of {\it c.d.f}. For larger samples, the equivalent outliers $\xi_i$ and $\xi_i\,\mathrm{mod}\,2\pi$  result in larger errors in comparison with the results provided by the interquantile distance method. 

Let us turn to the $\alpha$-dependence. Two methods perform slightly differently at small $\alpha$ (Fig.~\ref{compA}), otherwise reproducing the corresponding values very accurately. 
The root-mean-square deviation of $\hat{a}$ from the exact value is a monotonically decaying function of the L{\'e}vy index for the characteristic function method. 
For the interquartile method it has a minimum around $\alpha=1$: this fact reflects  
various sensitivity of the method for different $\alpha$; taking quartiles produces maximal sensitivity exactly for $\alpha=1$. 

\section{Practical example}
\label{Pe}

\begin{figure}
\includegraphics[width=\textwidth]{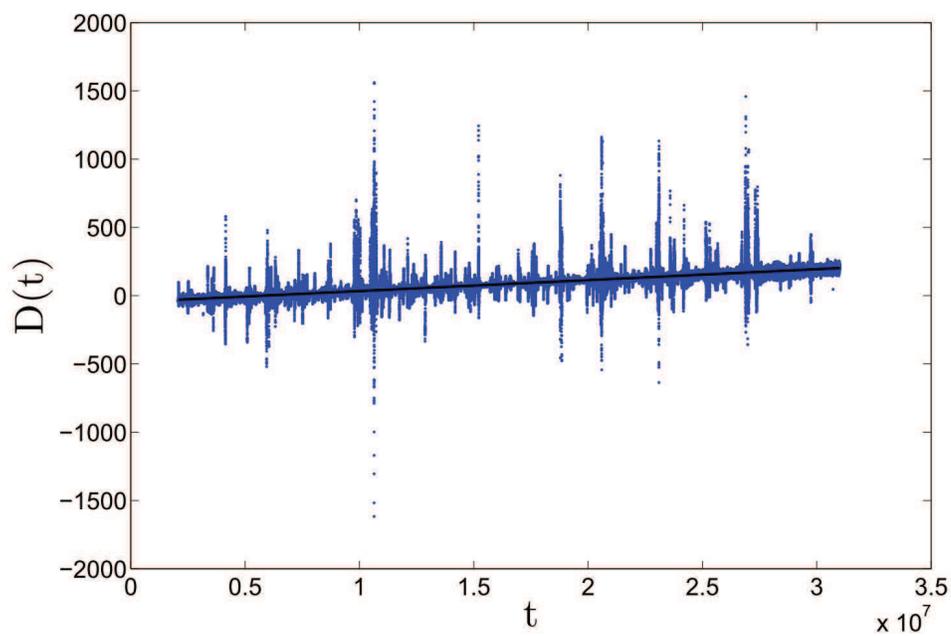}
\caption{Plot of the eastward component of the geomagnetic field D at Halley, Antarctica measured at X-min resolution from January 26 to December 28, 1998, after \cite{Clarke2003} (courtesy of M.P. Freeman, British Antarctic Survey) and the linear trend line with the coefficients determined via the scale parameter regression method. The time scale: seconds since January 1, 1998.}
\label{halley_data}
\end{figure}

As a practical test we process geomagnetic field data measured by a fluxgate magnetometer located at Halley, Antarctica on the Brunt ice shelf. Such data are known to  be complex comprising regular oscillations, 
highly irregular short bursts, and a linear trend originating from the ice shelf displacement \cite{Clarke2003}. It should be pointed out that the de-trending of such data is one of the key problems of ice shelf-based data processing \cite{Thomas1970}. 
Fig.~\ref{halley_data} demonstrates the example of such data, the small-scale processing of which has been discussed in the work \cite{Clarke2003}. 
Its authors highlighted the necessity of an additional median excluding even for very short portions of the data de-trended by a conventional method due to a presence of large outliers. 
The feature which makes this practical example different from our previous numerical ones is the correlated nature of the noise. 
However, one readily infers that the correlation time is short compared to the total measurement time, so that the methods should presumably work.  

The parameters of the noise were estimated as follows: the data were detrended by the least square fit, and then the routine {\tt stblfit} \cite{routine} 
was applied to check whether the de-trended distribution belongs to the class of alpha-stable ones. The process rapidly converges to the following parameters: 
the characteristic exponent $\alpha=1.39381$, the skewness $\beta=-0.0695959$, the scale parameter $\gamma=11.8844$ and the location $\delta=-2.33173$.
Thus, one can assume to a good approximation that (up to the correlated nature of the noise) the situation belongs to the class described above:
the practically symmetric L\'evy noise. The nonzero location parameter appears due to inconsistencies in determination of the shift parameter by the
usual least square approach as discussed below.

The estimates for $a$ and $b$ given by the least mean square (LMS) regression and by the scale parameter regression (SPR) are $(a,b)= \left(8.055\cdot10^{-6},-45.8\right)$
and $\left(8.006\cdot10^{-6},-46.4\right)$ respectively. The results of the quantile regression (QR) for different quantiles are given in Table \ref{intrqv}.

\begin{table} [h!]
\caption{The linear fit parameters for different interquantile distances.}
\begin{tabular}{cc}
Quantile interval&Parameters $(a,\,b)$ \\
\hline
$[0.25\,- \, 0.75]$&$(8.237\cdot10^{-6},-50.1)$\\
$[0.30\,- \,0.70]$&$(8.126\cdot10^{-6},-48.3)$\\
$[0.40\,- \,0.60]$&$(8.257\cdot10^{-6},-50.0)$\\
$[0.475\,- \,0.525]$&$(8.161\cdot10^{-6},-48.9)$\\
\end{tabular}
\label{intrqv}
\end{table}

\begin{figure}
\includegraphics[width=\textwidth]{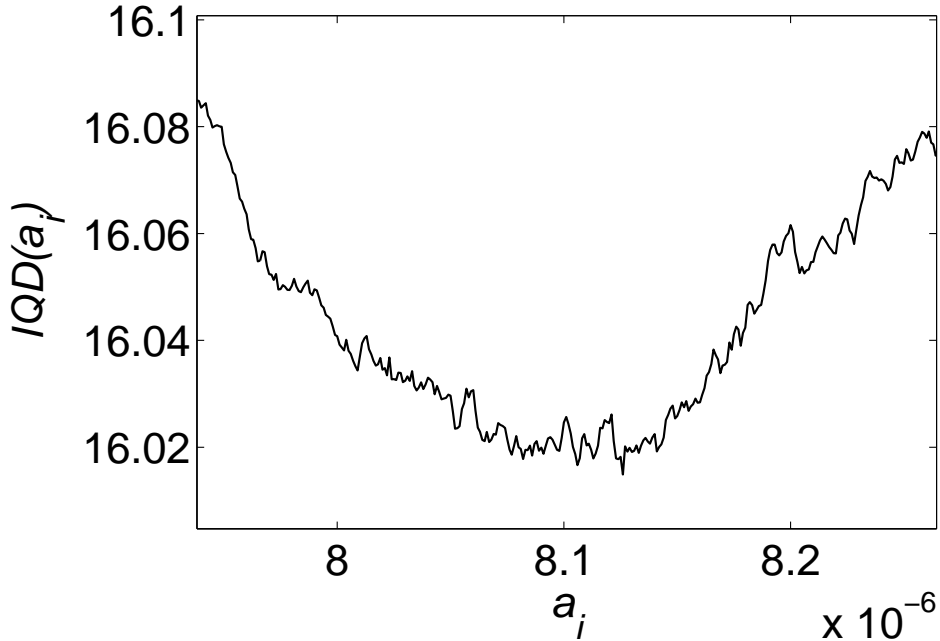}
\caption{The intequantile distance between $Q_{0.7}$ and $Q_{0.3}$ 
as a function of the trial slopes $a_i$ for the data shown in the Fig.~\ref{halley_data}.}
\label{HAmin}
\end{figure}

One readily infers that the results of application of LMS and SPR procedures are quite similar, while the results of QR are stable with respect to the choice of the quantile,
but overestimate the slope compared to the previous two methods. 
This fact can be traced back to the local irregularity of interquantile distance curve in the vicinity of a quite flat global minimum (see Fig.~\ref{HAmin}), whose flatness is partly 
due to the relatively large value of $\alpha$. Therefore, the fact that the interquantile regression, which \textit{on the average} might perform better than SPR for very long-tailed distributions 
in the case of small samples (compare with the results in Fig.~\ref{compA}), does not warrant for better performance in a single run for more regular noises and large samples.

Let us now concentrate on the comparison of SPR and LMS procedures as applied to the subdivisions of the whole sample and to the sample with excluded outliers, with the goal to compare new and conventional approach. 
The parameters of linear fits for a set of intervals obtained via the subdivision the initial time interval into two and four parts are presented in Table~\ref{tabsubd}. One readily infers that the relative 
variation of slopes does not exceed $9\%$ for the scale parameter regression in contrast to more then $20\%$ for the conventional least squares fit. The latter results even in more irregular behavior
of the shift parameter: for subdivision into four intervals it varies by a factor of 3 compared to merely $20\%$ as given by the robust method. Therefore, although large outliers influence the fitting results for both methods, 
the scale parameter regression allows for determination of the basic physical effect (speed of ice motion, which is a constant directly determining the trend's slope) more accurately. 

\begin{table}
\caption{The comparison of linear fit parameters $(k,\,b)$ obtained by two methods (Scale Parameter Regression -- SPR and Least Mean Square Regression -- LMS) for the subdivided time intervals expressed as a ratio to the whole interval taken as a unit.}

\begin{tabular}{ccc}
Subinterval&SPR&LMS\\
\hline
$[0,1]$&$\left(8.006\cdot10^{-6},-46.4\right)$&$\left(8.055\cdot10^{-6},-45.8\right)$\\
\hline
$[0,1/2]$&$\left(8.121\cdot10^{-6},-48.2\right)$&$\left(8.222\cdot10^{-6},-47.5\right)$\\
$[1/2,1]$&$\left(7.731\cdot10^{-6},-40.0\right)$&$\left(7.690\cdot10^6,-36.9\right)$\\
\hline
[0,1/4]&$\left(7.791\cdot10^{-6},-47.8\right)$&$\left(7.804\cdot10^{-6},-46.5\right)$\\
$[1/4,1/2]$&$\left(7.652\cdot10^{-6},-41.0\right)$&$\left(6.424\cdot10^{-6},-23.0\right)$\\
$[1/2,3/4]$&$\left(7.985\cdot10^{-6},-45.0\right)$&$\left(8.556\cdot10^{-6},-54.2\right)$\\
$[3/4,1]$&$\left(7.278\cdot10^{-6},-40.0\right)$&$\left(7.016\cdot10^{-6},-18.5\right)$\\
\hline
\end{tabular}
\label{tabsubd}
\end{table}

As the second test for comparison with standard approach to the processing of data with large outliers, we discuss linear fitting of the same sample with excluded  outliers. 
At the first step, we de-trended the data by the mean least square fit, stated the cutoff level, above which the points were excluded, and finally processed initial sample 
without the excluded points again.  

Table~\ref{outelim} shows the results of processing of regularized data in comparison with the original ones. As it should be, the exclusion of points, whose deviation exceed $1\%$ of the maximal detected value, 
results in the equal (within a prescribed accuracy) coefficients of the linear fit. While the results for SPR practically do not change when adding the points with larger deviations,
the ones of LMS show a considerable trend.

\begin{table}
\caption{The comparison of linear fit parameters $(k,\,b)$ obtained by two methods (Scale Parameter Regression -- SPR and Least Mean Square Regression -- LMS) after elimination of outliers.}

\begin{tabular}{ccc}
Cutoff level &SPR&LMS\\
\hline
100\%&$\left(8.006\cdot10^{-6},-46.4\right)$&$\left(8.055\cdot10^{-6},-45.8\right)$\\
25\%&$\left(8.007\cdot10^{-6},-46.4\right)$&$\left(8.034\cdot10^{-6},-46.2\right)$\\
1\%&$\left(8.008\cdot10^{-6},-46.5\right)$&$\left(8.007\cdot10^{-6},-46.8\right)$\\
\end{tabular}
\label{outelim}
\end{table}

\section{Conclusions}

The results of this work can be summarized as follows. We have discussed two methods for the robust linear fit to noisy signals, which can be applied to the case 
when the lower moments for the noise probability distribution diverge, e.g. for L{\'evy} noises. Both are based on the idea that the width 
of the distribution of the residues is the smallest when the slope of the regression line is chosen correctly, and differ in how this width is defined. 

The first method is the quantile regression approach. The second method deals with its counterpart in frequency domain, i.e. with the maximization of the trial characteristic function. 
Both approaches demonstrate their robustness and high accuracy for the noise distributions with extremely large outliers and may be used for a wide range of applications, 
for which such a behavior is characteristic, e.g. in problems of plasma dynamics, econophysics, etc. As a practical test we apply the methods to 
the data of the geomagnetic field measurements by a detector placed on an Antarctic ice shelf, showing large irregularity, and compare their performance to
the one of standard approaches. In this case the scale parameter regression seems to perform the best.

\section*{Acknowledgments}

We gratefully thank Dr. N.W. Watkins (LSE) for suggesting the geomagnetic data example and Dr. M.P. Freeman (British Antarctic Survey) for kindly providing the experimental data from Halley, Antarctica.

This work is partially supported by grant no. 1391 of the Ministry of Education and Science of the Russian Federation within the basic part of research funding no. 2014/349 assigned to Kursk State University
and by DFG (project SO 307/4-1).


\end{document}